\documentclass[twocolumn]{aastex631}

\usepackage{multirow}

\begin{document}

\title{A Mass Model for the Lensing Cluster SDSS~J1004+4112: Constraints From the Third Time Delay}

\author[0000-0002-6482-2180]{R. For\'es-Toribio}
\affiliation{Departamento de Astronom\'{\i}a y Astrof\'{\i}sica, Universidad de Valencia, E-46100 Burjassot, Valencia, Spain}
\affiliation{Observatorio Astron\'omico, Universidad de Valencia, E-46980 Paterna, Valencia, Spain}
\author[0000-0001-9833-2959]{J. A. Mu\~noz}
\affiliation{Departamento de Astronom\'{\i}a y Astrof\'{\i}sica, Universidad de Valencia, E-46100 Burjassot, Valencia, Spain}
\affiliation{Observatorio Astron\'omico, Universidad de Valencia, E-46980 Paterna, Valencia, Spain}
\author[0000-0001-6017-2961]{C. S. Kochanek}
\affiliation{Department of Astronomy, The Ohio State University, 140 West 18th Avenue, Columbus, OH 43210, USA}
\affiliation{Center for Cosmology and Astroparticle Physics, The Ohio State University, 191 West Woodruff Avenue, Columbus, OH 43210, USA}
\author[0000-0003-1989-6292]{E. Mediavilla}
\affiliation{Instituto de Astrof\'{\i}sica de Canarias, V\'{\i}a L\'actea S/N, La Laguna, E-38200, Tenerife, Spain}
\affiliation{Departamento de Astrof\'{\i}sica, Universidad de la Laguna, La Laguna, E-38200, Tenerife, Spain}

\correspondingauthor{Raquel For\'es-Toribio} \email{raquel.fores@uv.es}

\begin{abstract}

We have built a new model for the lens system SDSS~J1004+4112 including the recently measured time delay of the fourth quasar image. This time delay has a strong influence on the inner mass distribution of the lensing cluster ($\rho \propto r^{-\alpha}$) allowing us to determine $\alpha=1.18^{+0.02(+0.11)}_{-0.03(-0.18)}$ at the 68\% (95\%) confidence level in agreement with hydrodynamical simulations of massive galaxy clusters. We find an offset between the brightest cluster galaxy (BCG) and the dark matter halo of $3.8^{+0.6(+1.4)}_{-0.7(-1.3)}$ kpc at 68\% (95\%) confidence which is compatible with other galaxy cluster measurements. As an observational challenge, the estimated time delay between the leading image C and the faint (I=24.7) fifth image E is roughly 8 yr.

\end{abstract}

\section{Introduction} \label{sec:intro}

The lens system SDSS~J1004+4112 was the first example of a quasar lensed by a cluster \citep{2003Natur.426..810I}. The source and lens redshifts are $z_s=1.73$ and $z_l=0.68$, respectively. This leads to an image separation of 15\arcsec\ and large magnified images of the host galaxy. It is the second largest quasar lens after SDSS J1029+2623 \citep{2006ApJ...653L..97I}. Since its discovery, many additional observational constraints have been obtained: more galaxy cluster members \citep{2004ApJ...605...78O}, the central quasar image \citep{2005PASJ...57L...7I,2008PASJ...60L..27I}, other background lensed sources \citep{2005ApJ...629L..73S,2009MNRAS.397..341L,2010PASJ...62.1017O}, the velocity dispersion of the brightest galaxy cluster \citep{2008PASJ...60L..27I} and the time delays between quasar images A, B, and C \citep{2007ApJ...662...62F,2008ApJ...676..761F}.

As new observational constraints were measured, increasingly sophisticated mass models of the lens cluster were made using both parametric \citep{2003Natur.426..810I,2004ApJ...605...78O,2007ApJ...662...62F,2010PASJ...62.1017O} and nonparametric \citep{2004AJ....128.2631W,2006ApJ...652L...5S,2009MNRAS.397..341L,2015PASJ...67...21M} approaches. Recently, the time delay of the fourth quasar image \citep{Munoz_submitted} has been measured at 2457 days between images D and C, and it is over one year longer than predicted \citep{2010PASJ...62.1017O}. Indeed, the ability of the models to predict the unmeasured time delays has been remarkably poor.

Apart from the models of the mass distribution, this system has been widely studied not only in the optical but also from X-rays to radio \citep{2006A&A...454..493L,2006ApJ...647..215O,2009ApJ...702..472R,2011ApJ...739L..28J,2012ApJ...755...24C,2021MNRAS.505L..36M}. However, there are still issues that remain unsolved like the origin of the enhanced asymmetric wings in the broad emission lines of quasar image A \citep{2004ApJ...610..679R,2006ApJ...645L...5G,2006A&A...454..493L,2012ApJ...755...82M,2018ApJ...859...50F,2020A&A...634A..27P,2021A&A...653A.109F}.

Here we revisit the mass model making use of the newly measured time delay. The new delay should help to constrain the inner mass profile of the cluster halo and to look for deviations from a Navarro-Frenk-White (NFW) profile \citep{2006PASJ...58..271K,2010PASJ...62.1017O}. The paper is organized as follows. In Section \ref{sec:obs} we present the available observational constraints. In Section \ref{sec:model} we describe the modeling method and the components of the mass model. Section \ref{sec:results} presents the results and they are discussed in Section \ref{sec:disc}. Throughout this paper we assume a flat $\Lambda$CDM cosmology with $\Omega_M=0.26$, $\Omega_{\Lambda}=0.74$ and $H_0$=72 km s$^-1$ Mpc$^{-1}$.

\section{Observational Data \label{sec:obs}}

In addition to the lensed quasar, SDSS~J1004+4112 has seven lensed background galaxies at three different redshifts. Their redshifts, image positions, and magnitudes (with respect to quasar image A) are from \citet{2010PASJ...62.1017O} and listed in Table \ref{tab:images}. For the quasar time delays we use the results of \citet{Munoz_submitted} with the uncertainties set to five times the formal errors (see Table \ref{tab:images}) since time delay uncertainties are almost always underestimated \citep[see, e.g.,][]{2018MNRAS.473...80T}. We also adopted the positions, ellipticities, position angles, and luminosity ratios (with respect to the central galaxy) of 14 cluster galaxy members (see Table \ref{tab:satgal}) from \citet{2010PASJ...62.1017O}. For the brightest cluster galaxy (BCG) we adopt a position of (7\farcs114,4\farcs409), an ellipticity of $e=0.30\pm0.05$, a major axis PA of $\theta_e=152\pm5$\degr\ \citep{2010PASJ...62.1017O} and a central velocity dispersion of $352\pm13$ km s$^-1$ \citep{2008PASJ...60L..27I}.
\begin{deluxetable}{ccDDr@{$\pm$}lr@{$\pm$}l}
\tabletypesize{\footnotesize}
\tablecaption{Redshift, Positions, Magnitudes, and Time Delays for the Observed Images of the Lensed Sources\label{tab:images}}
\tablehead{
\colhead{Name} & \colhead{$z_s$} & \multicolumn2c{$x$ (\arcsec)} & \multicolumn2c{$y$ (\arcsec)} & \multicolumn2c{$\Delta m$} & \multicolumn2c{$\Delta t$ (days)}
}
\decimals
\startdata
QSO.A &  & 0.000 & 0.000 & \multicolumn2c{$\equiv$0} & 825.99&2.10 \\
QSO.B &  & $-$1.317 & 3.532 & 0.35&0.30 & 781.92&2.20 \\
QSO.C & 1.734 & 11.039 & $-$4.492 & 0.87&0.30 & \multicolumn2c{$\equiv$0} \\
QSO.D &  & 8.399 & 9.707 & 1.50&0.30 & 2456.99&5.55 \\ 
QSO.E &  & 7.197 & 4.603 & 6.3&0.8 & \multicolumn2c{\nodata} \\
\hline
A1.1 &  & 3.93 & $-$2.78 & \multicolumn2c{\nodata} & \multicolumn2c{\nodata} \\
A1.2 &  & 1.33 & 19.37 & \multicolumn2c{\nodata} & \multicolumn2c{\nodata} \\
A1.3 & 3.33 & 19.23 & 14.67 & \multicolumn2c{\nodata} & \multicolumn2c{\nodata} \\
A1.4 &  & 18.83 & 15.87 & \multicolumn2c{\nodata} & \multicolumn2c{\nodata} \\
A1.5 &  & 6.83 & 3.22 & \multicolumn2c{\nodata} & \multicolumn2c{\nodata} \\
\hline
A2.1 &  & 4.13 & $-$2.68 & \multicolumn2c{\nodata} & \multicolumn2c{\nodata} \\
A2.2 &  & 1.93 & 19.87 & \multicolumn2c{\nodata} & \multicolumn2c{\nodata} \\
A2.3 & 3.33 & 19.43 & 14.02 & \multicolumn2c{\nodata} & \multicolumn2c{\nodata} \\
A2.4 &  & 18.33 & 15.72 & \multicolumn2c{\nodata} & \multicolumn2c{\nodata} \\
A2.5 &  & 6.83 & 3.12 & \multicolumn2c{\nodata} & \multicolumn2c{\nodata} \\
\hline
A3.1 &  & 4.33 & $-$1.98 & \multicolumn2c{\nodata} & \multicolumn2c{\nodata} \\
A3.2 &  & 2.73 & 20.37 & \multicolumn2c{\nodata} & \multicolumn2c{\nodata} \\
A3.3 & 3.33 & 19.95 & 13.04 & \multicolumn2c{\nodata} & \multicolumn2c{\nodata} \\
A3.4 &  & 18.03 & 15.87 & \multicolumn2c{\nodata} & \multicolumn2c{\nodata} \\
A3.5 &  & 6.83 & 3.02 & \multicolumn2c{\nodata} & \multicolumn2c{\nodata} \\
\hline
B1.1 &  & 8.88 & $-$2.16 & \multicolumn2c{\nodata} & \multicolumn2c{\nodata} \\
B1.2 & 2.74 & $-$5.45 & 15.84 & \multicolumn2c{\nodata} & \multicolumn2c{\nodata} \\
B1.3 &  & 8.33 & 2.57 & \multicolumn2c{\nodata} & \multicolumn2c{\nodata} \\
\hline
B2.1 &  & 8.45 & $-$2.26 & \multicolumn2c{\nodata} & \multicolumn2c{\nodata} \\
B2.2 & 2.74 & $-$5.07 & 16.04 & \multicolumn2c{\nodata} & \multicolumn2c{\nodata} \\
B2.3 &  & 8.33 & 2.57 & \multicolumn2c{\nodata} & \multicolumn2c{\nodata} \\
\hline
C1.1 &  & 10.25 & $-$3.06 & \multicolumn2c{\nodata} & \multicolumn2c{\nodata} \\
C1.2 & 3.28 & $-$7.55 & 15.39 & \multicolumn2c{\nodata} & \multicolumn2c{\nodata} \\
C1.3 &  & 8.49 & 2.72 & \multicolumn2c{\nodata} & \multicolumn2c{\nodata} \\
\hline
C2.1 &  & 9.95 & $-$3.36 & \multicolumn2c{\nodata} & \multicolumn2c{\nodata} \\
C2.2 & 3.28 & $-$7.30 & 15.44 & \multicolumn2c{\nodata} & \multicolumn2c{\nodata} \\
C2.3 &  & 8.49 & 2.72 & \multicolumn2c{\nodata} & \multicolumn2c{\nodata} \\
\enddata
\tablecomments{The position errors adopted for the QSO are 0\farcs04 whereas for the galaxies are 0\farcs4.}
\end{deluxetable}
\begin{deluxetable}{DDDDD}
\tablecaption{Positions, Ellipticities, Position Angles, and Luminosity Ratios of 14 Cluster Galaxy Members\label{tab:satgal}}
\tablehead{
\multicolumn2c{$x$ (\arcsec)} & \multicolumn2c{$y$ (\arcsec)} & \multicolumn2c{e} & \multicolumn2c{PA (\degr)} & \multicolumn2c{$L/L_{BCG}$}
}
\decimals
\startdata
30.78 & 4.50 & 0.2723 & $-$2.90 & 0.5050 \\
12.14 & 3.67 & 0.2426 & $-$131.80 & 0.2970 \\
2.76 & 14.13 & 0.1077 & $-$161.50 & 0.2250 \\
25.29 & $-$9.06 & 0.1028 & $-$70.00 & 0.0950 \\
$-$9.22 & $-$2.53 & 0.0885 & $-$42.80 & 0.1510 \\
14.54 & 24.23 & 0.0461 & $-$82.00 & 0.1830 \\
24.61 & 4.72 & 0.0433 & $-$60.40 & 0.1720 \\
9.36 & 2.41 & 0.2249 & $-$100.30 & 0.4060 \\
2.767 & $-$0.171 & 0.0517 & $-$21.8 & 0.238 \\
14.789 & $-$5.454 & 0.0453 & $-$127.7 & 0.41 \\
$-$1.359 & 0.482 & 0.0224 & 51.3 & 0.21 \\
12.00 & 13.82 & 0.0711 & $-$92.7 & 0.368 \\
7.84 & 9.10 & 0.0118 & $-$17.0 & 0.10 \\
$-$7.21 & $-$8.84 & 0.0661 & $-$139.6 & 0.450 \\
\enddata
\end{deluxetable}

\section{System Modeling} \label{sec:model}

We modeled the mass distribution of the lens using both \textit{lensmodel} \citep{2001astro.ph..2340K} and \textit{glafic} \citep{2010PASJ...62.1017O}. Both packages employ parametrized mass profiles and optimize the model using the downhill simplex method \citep[see, e.g.,][]{1992nrca.book.....P}. The final results are computed with \textit{glafic} in order to ease the comparison with \citet{2010PASJ...62.1017O}. The different mass components and any constraints on their properties are described in the following subsections.

\subsection{Dark Matter Halo}

We modeled the dark matter (DM) mass distribution as a generalized Navarro-Frenk-White (gNFW) profile \citep{1997ApJ...490..493N,2000ApJ...529L..69J} with a 3D density profile
\begin{equation}\label{eq:gnfw}
    \rho=\frac{\rho_s}{(r/r_s)^{\alpha}(1+r/r_s)^{3-\alpha}}.
\end{equation}
With the addition of its 2D ellipticity, the projected profile is completely described by seven parameters: $M_{vir}$ (virial mass), $x$ and $y$ (position), e (ellipticity), $\theta_e$ (position angle), $c_{-2}$ (concentration parameter), and $\alpha$ (inner slope). The relationships between ($\rho_r,r_s$) and ($M_{vir},c_{-2}$) can be found in \textit{glafic}'s manual\footnote{Available at \url{https://www.slac.stanford.edu/~oguri/glafic/index_v1.html}}. We do not impose any restrictions on the parameters, however the inner slope $\alpha$ is set to unity until the whole model is optimized.

\subsection{BCG}

The BCG is parametrized by a pseudo-Jaffe (pJaffe) profile
\begin{equation}
    \kappa=2\pi\left(\frac{\sigma}{c}\right)^2\frac{D_{ls}}{D_{os}} \left[ \frac{1}{R}-\frac{1}{\sqrt{r_{trun}^2+R^2}} \right].
\end{equation}
This model is defined by six parameters when ellipticity is added: $\sigma$ (velocity dispersion), $x$ and $y$ (position), e (ellipticity), $\theta_e$ (position angle) and $r_{trun}$ (truncation radius, where beyond this scale, the convergence falls as $R^{-3}$). We fix the position to the observed value and include Gaussian priors on e and $\theta_e$. The observed velocity dispersion does not directly correspond to the model dispersion and the Appendix \ref{sec:vdisp} details how we use the measured dispersion as a constraint to obtain a model velocity dispersion prior of 325$\pm$20 km s$^-1$. The truncation radius is loosely constrained using the same prior as \citet{2010PASJ...62.1017O} ($r_{trun}=8\arcsec\pm4\arcsec$) because of the observed correlation between the velocity dispersion and truncation radius \citep{2009ApJ...693..970N}.

\subsection{Other Cluster Galaxy Members}

The rest of the cluster galaxies are modeled by scaled pseudo-Jaffe ellipsoids (mass profile `gals' in the \textit{glafic} software). Their $\sigma$ and $r_{trun}$ are scaled relative to the luminosity of the BCG as
\begin{equation}
    \sigma=\sigma_* \left( \frac{L}{L_*}\right)^{1/4} \;  \textnormal{ and } \; r_{trun}=r_*\left( \frac{L}{L_*}\right)^{1/2}
\end{equation}
\noindent where $\sigma_*$ and $r_*$ are model parameters. The ratio $L/L_*$, ellipticity, and position angle of each galaxy member are constrained by the observed values given in Table \ref{tab:satgal}. Since the reference galaxy is the BCG, the scale parameters $\sigma_*$ and $r_*$ are initially set equal to the velocity dispersion and truncation radius of the BCG, but, during the optimization, both are allowed to vary.

\vspace{0.1cm}
\subsection{External Perturbations}

We also include multipole perturbations to mimic deviations of the DM  halo from a perfect ellipsoid. Like \citet{2010PASJ...62.1017O} we used multipoles of order m=2, 3, 4, and 5 (mpoles model) with the potential
\begin{equation}
    \phi=-\frac{\epsilon}{m}r^2\cos m (\theta-\theta_{\epsilon}-\pi/2).
\end{equation}
For m=2, the perturbation is an external shear (named `pert' in \textit{glafic}) so we denote the m=2 parameters as $\gamma$ and $\theta_{\gamma}$ instead. We center these potentials on the position of the DM halo.

\subsection{Background Sources}

At higher redshifts than the quasar there are three groups of galaxies (A, B, and C) at different redshifts. Group A is composed of three galaxies each lensed into five images. Groups B and C each have two galaxies and each is split into three images. The observed positions of these images along with their redshifts and magnitudes are also given in Table \ref{tab:images}. All the background sources are modeled as point sources with fixed redshifts so the fit parameters are their unlensed positions.\\

In total we have 77 observational constraints and 37 model parameters leaving us with $\nu$=40 degrees of freedom.

\section{Results}
\label{sec:results}

The optimization was performed as follows. First we optimized all the parameters fitting the images on the source plane. Once we found the minimum $\chi^2$, we ran the optimization fitting the images on the image plane. Next we computed the parameter uncertainties fitting the images on the source plane in order to maintain a sensible computing time. Finally, we ran the optimization on the image plane using as initial parameters the model that produced the minimum $\chi^2$ in the uncertainties estimation step.

\begin{figure*}
\includegraphics[width=\linewidth]{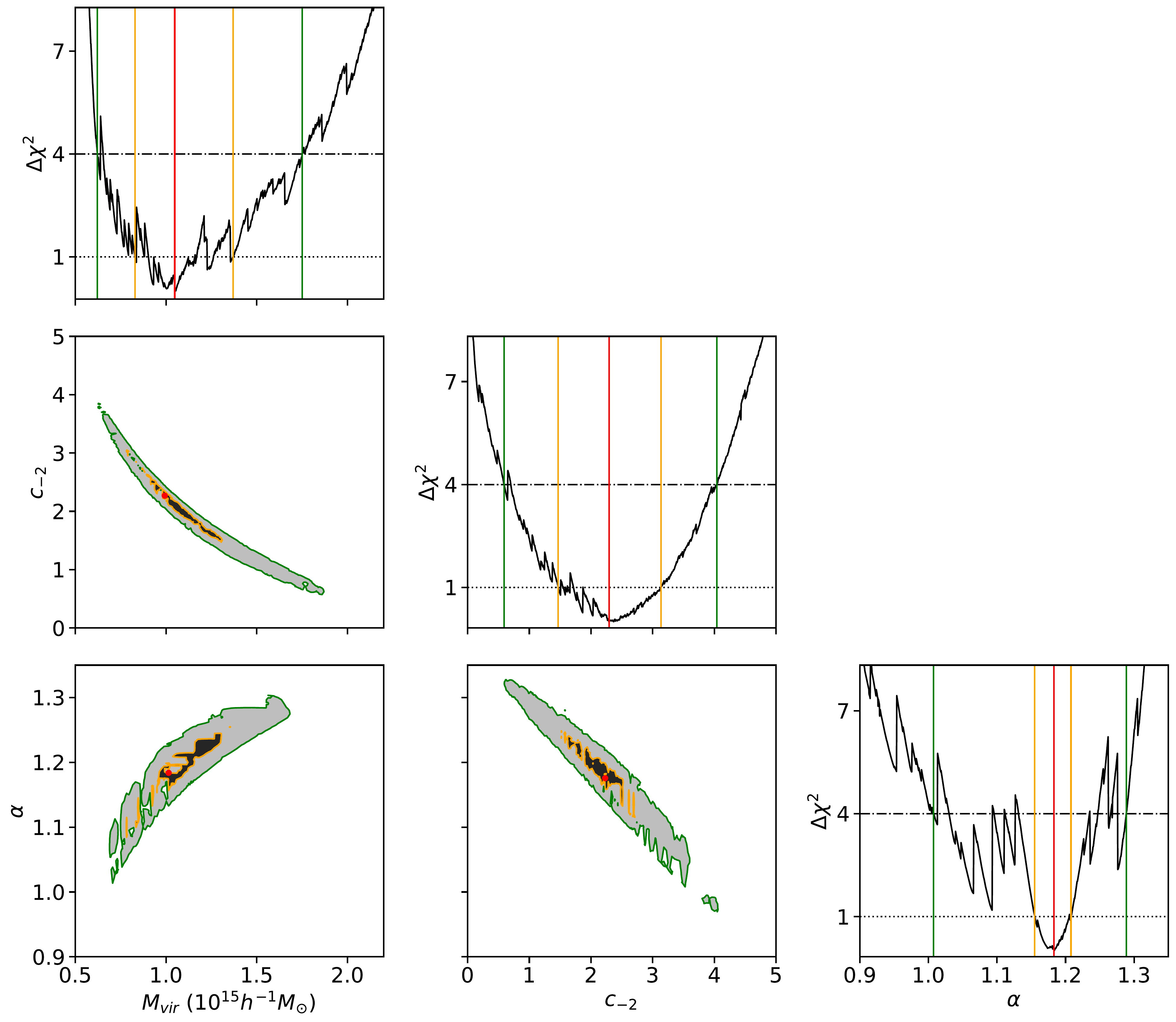}
\caption{$\chi^2$ contours and profiles for the virial mass, concentration, and inner slope parameters of the generalized NFW model (Eqn. \ref{eq:gnfw}). Red lines and points mark the values for which the $\chi^2$ is minimized, orange lines delimiter the 1$\sigma$ contours, and green lines the 2$\sigma$ contours. The black (gray) shadow is the 1$\sigma$ (2$\sigma$) region.
\label{fig:chicut}}
\end{figure*}

The central values and the 1$\sigma$ and 2$\sigma$ (corresponding to $\Delta \chi^2=1$ and $\Delta \chi^2=4$, respectively) uncertainties are given in Table \ref{tab:modparam}. We define the uncertainties using the model most distant from the $\chi^2$ minimum that is within the $\Delta\chi^2$ limit. The $\chi^2$ contours and profiles for the mass $M_{vir}$, concentration $c_{-2}$ and inner slope $\alpha$ of the gNFW model are displayed in Figure \ref{fig:chicut}. The parameters $M_{vir}$ and $c_{-2}$ are strongly correlated because they are defined on scales much larger than the Einstein ring, an issue we discuss in more detail below. The parameters do not have smooth $\chi^2$ profiles because the minimizer has difficulties in finding the true minimum.
\begin{table}
\centering
\caption{Central Values and Errors at 1-$\sigma$ and 2-$\sigma$ for the Mass Model Parameters\label{tab:modparam}}
\begin{tabular}{ccccc|c}
\hline
\hline
Model & Parameters &  & 1$\sigma$ & 2$\sigma$ & Best Fit\\
\hline
\multirow{7}{*}{gNFW} & $M_{vir}$ ($10^{15} h^{-1} M_{\sun}$) & 1.0 & $^{+0.3}_{-0.2}$ & $ ^{+0.7}_{-0.4}$ & 1.0 \\
    & $x$ (\arcsec) & 7.04 & $ ^{+0.07}_{-0.07}$ & $ ^{+0.13}_{-0.14}$ & 7.03\\
    & $y$ (\arcsec) & 4.95 & $ ^{+0.09}_{-0.10}$ & $ ^{+0.2}_{-0.19}$ & 4.88\\
    & e & 0.36 & $ ^{+0.02}_{-0.03}$ & $ ^{+0.04}_{-0.05}$ & 0.36\\
    & $\theta_e$ (\degr) & 158.6 & $ ^{+1.5}_{-1.8}$ & $ ^{+3}_{-3}$ & 158.1\\
    & $c_{-2}$ & 2.3 & $ ^{+0.8}_{-0.8}$ & $ ^{+1.7}_{-1.7}$ & 2.2\\
    & $\alpha$ & 1.18 & $ ^{+0.02}_{-0.03}$ & $ ^{+0.11}_{-0.18}$ & 1.20\\
\hline
\multirow{4}{*}{pJaffe} & $\sigma$ (km s$^-1$) & 296 & $ ^{+5}_{-4}$ & $ ^{+10}_{-12}$ & 290\\
    & e & 0.40 & $ ^{+0.06}_{-0.04}$ & $ ^{+0.12}_{-0.10}$ & 0.37\\
    & $\theta_e$ (\degr) & 152 & $ ^{+4}_{-3}$ & $ ^{+9}_{-7}$ & 154\\
    & $r_{trun}$ (\arcsec) & 10 & $ ^{+2}_{-2}$ & $ ^{+6}_{-4}$ & 10\\
\hline
\multirow{2}{*}{gals} & $\sigma_*$ (km s$^-1$) & 351 & $ ^{+12}_{-10}$ & $ ^{+20}_{-20}$ & 337\\
    & $r_*$ (\arcsec) & 9 & $ ^{+3}_{-2}$ & $ ^{+6}_{-4}$ & 9\\
\hline
\multirow{2}{*}{pert} & $\gamma$ ($10^{-2}$) & 5.8 & $^{+1.0}_{-1.2}$ & $^{+2}_{-2}$ & 5.7\\
    & $\theta_{\gamma}$ (\degr) & 66 & $ ^{+3}_{-5}$ & $ ^{+7}_{-9}$ & 65\\
\hline
\multirow{2}{*}{mpole3} & $\epsilon$ ($10^{-2}$) & 1.6 & $ ^{+0.3}_{-0.3}$ & $ ^{+0.6}_{-0.6}$ & 1.4\\
    & $\theta_{\epsilon}$ (\degr) & -9 & $ ^{+2}_{-3}$ & $ ^{+6}_{-7}$ & -7\\
\hline
\multirow{2}{*}{mpole4} & $\epsilon$ ($10^{-2}$) & 1.1 & $ ^{+0.2}_{-0.2}$ & $ ^{+0.5}_{-0.4}$ & 1.1\\
    & $\theta_{\epsilon}$ (\degr) & 40 & $ ^{+3}_{-4}$ & $ ^{+8}_{-7}$ & 41\\
\hline
\multirow{2}{*}{mpole5} & $\epsilon$ ($10^{-2}$) & 1.39 & $ ^{+0.13}_{-0.2}$ & $ ^{+0.3}_{-0.3}$ & 1.25\\
    & $\theta_{\epsilon}$ (\degr) & 16.2 & $ ^{+2}_{-1.8}$ & $ ^{+6}_{-4}$ & 17.6\\
\hline
\end{tabular}
\end{table}

The optimization on the image plane starting from the parameters which give the minimum $\chi^2$ in the error computations is considered the best-fit model and it is shown in the last column of Table \ref{tab:modparam}. The central values of the $\chi^2$ profiles do not exactly correspond to the best-fit parameters (see columns 3 and 6 of Table \ref{tab:modparam}) because the two results come from different optimization procedures. Nonetheless, all the parameters of the best model fit lie between the reported 1$\sigma$ confidence intervals except for $\sigma$ and $\sigma_*$ which lie between their 2$\sigma$ intervals.

\begin{figure*}
\includegraphics[width=\linewidth]{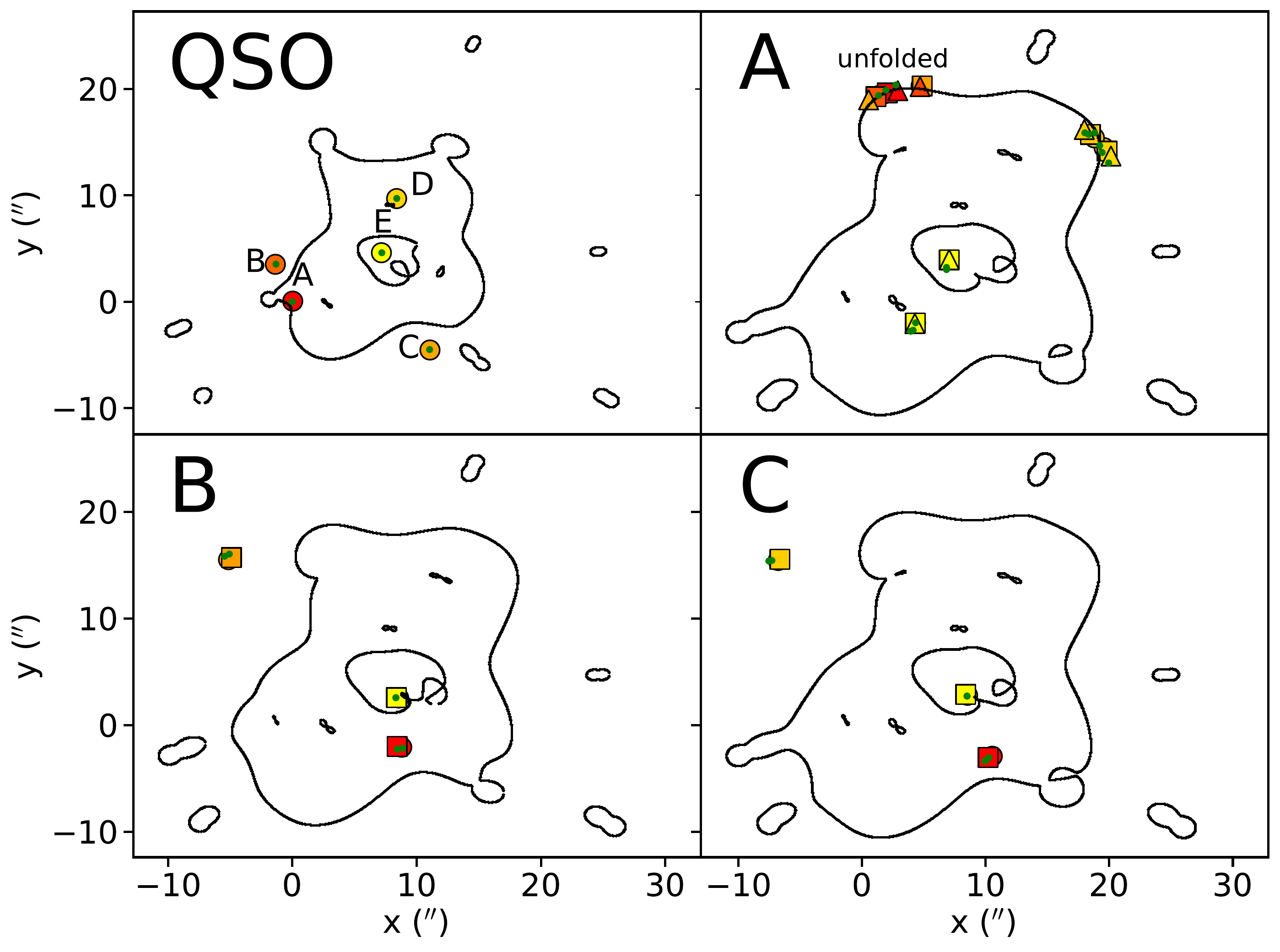}
\caption{Critical curves and images positions for the four background sources. The green dots mark the observed positions of the images. Circles denote the quasar, A1, B1, and C1 images, squares represent the images of A2, B2, and C2, and lastly the triangles mark the A3 images. Additionally, images are color coded depending on their flux, from the brightest in red to the dimmest in yellow.
\label{fig:imcrit}}
\end{figure*}

The predicted image positions and critical curves of the four background sources for this model are shown in Figure \ref{fig:imcrit}. This best-fit model has a $\chi^2$ value of 59.91 in the image plane which corresponds to $\chi^2_{\nu}=\chi^2/\nu=1.50$. The observational constraints on the quasar contribute 11.76 to the total $\chi^2$ and the A, B, and C groups of galaxies contribute 30.95, 2.92, and 8.16, respectively. Additionally, the $\chi^2$ associated with the six parameter priors is 6.13. The model fits all three delays well, with a $\chi^2$=1.33 for the delay constraints. The model predicts a time delay between the leading image C and the fifth image E of $\Delta t_{EC}$=2853.90 days or 7.81 yr, although this delay will be virtually impossible to measure due to the faintness of image E. Additionally, Table \ref{tab:kappagamma} gives the convergence $\kappa$ and shear $\gamma$ at the quasar positions along with the total magnification.
\begin{deluxetable}{cDDD}
\tablecaption{Convergence $\kappa$, Shear $\gamma$ and Magnification $\mu$ at the Quasar Image Positions\label{tab:kappagamma}}
\tablehead{
\colhead{Quasar Image} & \multicolumn2c{$\kappa$} & \multicolumn2c{$\gamma$} & \multicolumn2c{$\mu$}
}
\decimals
\startdata
A & 0.728940 & 0.333242 & $-$26.612171 \\
B & 0.650724 & 0.232737 & 14.743285 \\
C & 0.587686 & 0.225075 & 8.379131 \\
D & 1.015191 & 0.489943 & $-$4.169912 \\
E & 5.669724 & 3.622670 & 0.115173 \\
\enddata
\end{deluxetable}

\begin{figure*}
\gridline{\includegraphics[width=0.49\linewidth]{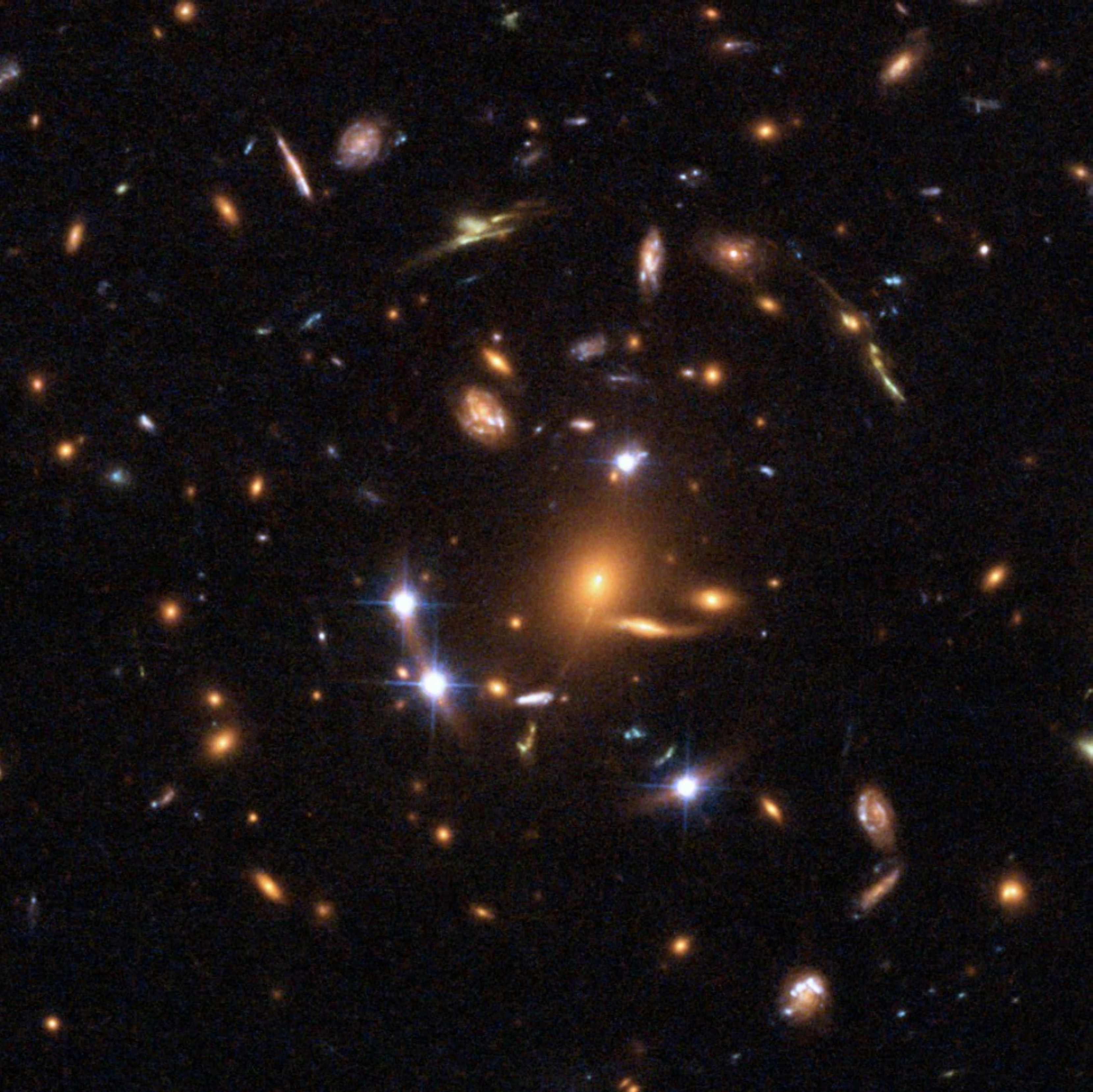}
          \includegraphics[width=0.49\linewidth]{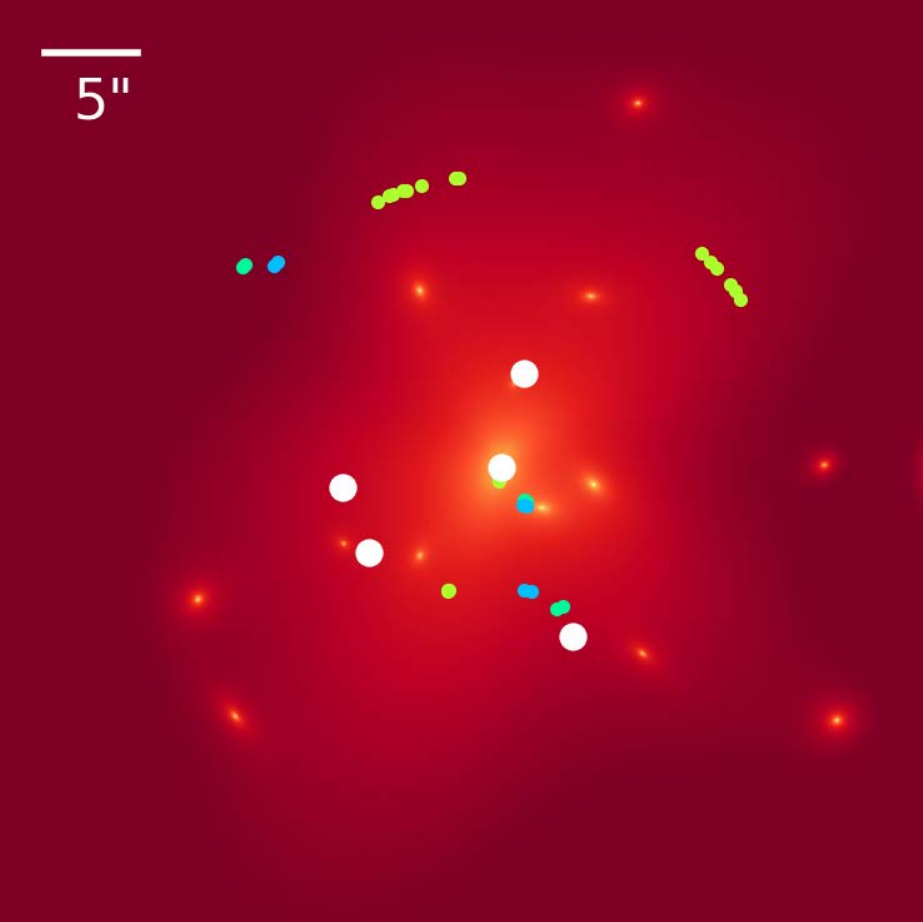}
          }
\caption{The left panel shows the Hubble Space Telescope Advanced Camera for Surveys (ACS)/WFC (GO-9744, PI: C. Kochanek) images of the lens. The right panel shows the reconstructed mass distribution on a logarithmic scale along with the images positions: quasar in white, and galaxy group A in lime-green, B in blue, and C in turquoise. The orientation for both figures is north up, west right. 
\label{fig:obsmodsys}}
\end{figure*}

The goodness of the fit for the A group of galaxies is substantially worse than for the rest of the background galaxies because the model predicts seven images instead of the five observationally reported. This occurs because the images A1.2, A2.2, and A3.2 lie very close to a critical curve. A very small change in the source position can change the number of the images and that is precisely what occurred in our model where one image has unfolded into three (they are marked as `unfolded' in Figure \ref{fig:imcrit}). The A group images are really composed of extended arcs so the problem would be solved by increasing their position uncertainties, although the model would be more realistic if we modeled the galaxies as extended sources. Despite that, the reconstructed image positions properly reproduce all the main structures of the system (see Figure \ref{fig:obsmodsys}).

\section{Discussion and Conclusions}
\label{sec:disc}

One of the main results from our model is the value of the inner slope of the generalized NFW profile, $\alpha=1.18^{+0.02(+0.11)}_{-0.03(-0.18)}$ at the 68\% (95\%) confidence level. We must bear in mind that this value is the asymptotic slope of the DM for $r \rightarrow 0$. However at this regime the mass of the BCG becomes important and it is difficult to distinguish both components because the lensing effect is sensitive to the total mass of the system. A different parameterization of the BCG would alter the inner slope value of the generalized NFW profile. If we wanted to determine the mass distribution in the innermost regions of the cluster more precisely, we would probably need a velocity dispersion profile in order to separate the two components better.

\citet{2013ApJ...765...24N,2013ApJ...765...25N} made such models for seven massive galaxy clusters and found very shallow inner slopes of $0.50\pm0.10\textnormal{(random)}^{+0.14}_{-0.13}\textnormal{(systematic)}$. On the other hand, \citet{2017MNRAS.471..383S} inferred a DM inner slope for the cluster Abell 1201 of $\alpha=1.01\pm0.12$ using the same procedure but with other assumptions. \citet{2015MNRAS.452..343S} and \citet{2020MNRAS.496.4717H} tested this method with the EAGLE hydrodynamical simulations and found that the results were dependent on the assumptions about the mass-to-light ratio and the intrinsic degeneracies of the profiles. The simulations suggest that the DM halo distribution is compatible with the original NFW profile and our value. While our model likely has degeneracies between the inner slope of the BCG (which we hold fixed) and the DM halo,
our result is compatible with these simulations of massive galaxy clusters.

Our inferred virial mass and concentration parameter for the generalized NFW cluster halo are in good agreement with those obtained with the lens model of \citet{2010PASJ...62.1017O}. However they differ from the values reported by \citet{2006ApJ...647..215O} based on X-ray observations, $M_{vir}=4.2^{+2.6}_{-1.5} \times 10^{14} \; h^{-1}M_{\sun}$ and $c_{-2}=6.1^{+1.5}_{-1.2}$. These differences can be explained by the degeneracy between $M_{vir}$, $c_{-2}$ and $\alpha$ in strong lensing models pointed out by \citet{2010PASJ...62.1017O} and shown for our model in Figure \ref{fig:chicut}. The source of this degeneracy comes from the fact that the lensing effect is mainly sensitive to the enclosed mass projected inside the region delimited by the lensed images but the parameters $M_{vir}$ and $c_{-2}$ are defined on the much larger scale of the virial radius and so are extrapolations of our mass model. In this system, the outermost images are $\sim$125 kpc from the center of the cluster whereas the parameters $M_{vir}$ and $c_{-2}$ are extrapolations to a radius $r_{vir} \approx$ 2 Mpc. It is better to test the compatibility of the two results by comparing the mass enclosed inside a smaller radius to which both methods are sensitive. \citet{2006ApJ...647..215O} reported the projected mass inside 70 h$^{-1}$kpc to be $3.5^{+1.3}_{-0.8} \times 10^{13} \; h^{-1}M_{\sun}$, in good agreement with our enclosed mass of $3.7 \times 10^{13} \; h^{-1}M_{\sun}$.

We find an offset between the BCG and the DM halo of $3.8^{+0.6(+1.4)}_{-0.7(-1.3)}$ kpc at the 68\% (95\%) confidence level. Previous parametric models have found similar offsets for this system \citep{2004ApJ...605...78O,2007ApJ...662...62F} although \citet{2010PASJ...62.1017O} found an offset compatible with zero. From fitting 10,000 galaxy clusters  with strong lensing measurements, \citet{2012MNRAS.426.2944Z} found a mean separation of $18^{+37}_{-12}$ kpc where the uncertainty is the observed scatter. \citet{2020ApJS..247...43K} estimated the separation between the BCG and the intracluster light tracing the DM distribution for local clusters, and found a mean offset of 36 kpc with a sample scatter of 33 kpc. Our inferred offset is compatible with both results.

We also find that the DM halo and the BCG are virtually aligned and with similar ellipticities. On the other hand, the velocity dispersion scale parameter for the cluster galaxy members differs from the BCG parameterization even though they were started with values scaled to the BCG. This difference suggests that BCGs are structurally different from the other cluster galaxy members because they have more DM content than the rest. Another feature in the BCG velocity dispersion is that the model fits the pseudo-Jaffe velocity dispersion to a smaller value than the prior. That could be explained by the fact that the observed velocity dispersion includes the effect of the DM halo whereas the model velocity dispersion parameter represents only the mass of the BCG.

In conclusion, we have made use of the recently measured time delay of image D and previous observational data to model the lensing cluster SDSS~J1004+4112. The model parameters we obtained are broadly consistent with previous estimates but they are now better constrained, in particular the inner slope of the DM halo of the lensing cluster.

\begin{acknowledgments}
This research is based on observations made with the NASA/ESA Hubble Space Telescope obtained from the Space Telescope Science Institute, which is operated by the Association of Universities for Research in Astronomy, Inc., under NASA contract NAS 5–26555. These observations are associated with program GO-9744. J.A.M. and E.M. are supported by the Spanish Ministerio de Ciencia e Innovación with the grants PID2020-118687GB-C32 and PID2020-118687GB-C31. J.A.M. is also supported by the Generalitat Valenciana with the project of excellence Prometeo/2020/085. C.S.K. is supported by NSF grants AST-1908570 and AST-1814440. 
\end{acknowledgments}

\appendix

\section{Model and Observed Velocity Dispersions}
\label{sec:vdisp}

The \textit{glafic} package parametrizes the pseudo-Jaffe profile with a velocity dispersion which corresponds to the central velocity dispersion of a singular isothermal sphere ($\sigma_{mod}$). On the other hand, the observed velocity dispersion is the averaged velocity dispersion ($\sigma_{obs}$) along the line of sight inside a finite size slit. The observed velocity dispersion is produced by a combination of the BCG mass and the DM halo, however we assume that the DM halo is negligible for a central velocity dispersion and we only consider the dispersion velocity due to the BCG modeled as a pseudo-Jaffe ellipsoid. In order to obtain their relationship we make use of the expressions of the densities, masses, and velocities dispersion of \citet{2007arXiv0710.5636E}. Additionally, we transform the expressions into dimensionless quantities plus a prefactor in order to simplify the expressions and we end with the following definitions:

\begin{itemize}

    \item{Surface density:} 
    \begin{equation}
        \Sigma(R)=\frac{\sigma_{mod}^2}{2Gr_{trun}} \tilde{\Sigma}(x) \;  \textnormal{ where } \; \tilde{\Sigma}(x)=\frac{1}{\sqrt{f^2+x^2}}-\frac{1}{\sqrt{1+x^2}}
    \end{equation}
    
    \item{Volume density:} 
    \begin{equation}
        \rho(r)=\frac{\sqrt{q}\left(1-f^2\right)\sigma_{mod}^2}{2 \pi G r^2_{trun}} \tilde{\rho}(z) \;  \textnormal{ where } \; \tilde{\rho}(z)=\frac{1}{\left(f^2+z^2\right)\left(1+z^2\right)}
    \end{equation}
    
    \item{Enclosed projected mass:}
    \begin{equation}
        M_{2D}(R)=\frac{\pi r_{trun} \sigma_{mod}^2}{G q} \tilde{M}_{2D}(x) \;  \textnormal{ where } \; \tilde{M}_{2D}(x)=\sqrt{f^2+x^2}-\sqrt{1+x^2}+1-f
    \end{equation}

    \item{Enclosed mass:}
    \begin{equation}
        M_{3D}(r)=\frac{2 r_{trun} \sigma_{mod}^2}{G q} \tilde{M}_{3D}(z) \;  \textnormal{ where } \; \tilde{M}_{3D}(z)=\arctan{\left(z\right)}-f\arctan{\left(z/f\right)}
    \end{equation}

    \item{Projected velocity dispersion:}
    \begin{equation}
        \sigma_P^2(R)=\frac{4 \left(1-f^2\right) \sigma_{mod}^2}{\pi \sqrt{q}} \tilde{\sigma_P^2}(x) \;  \textnormal{ where } \; \tilde{\sigma_P^2}(x)=\frac{1}{\tilde{\Sigma}(x)}\int_x^{\infty}\frac{\tilde{M}_{3D}(z)\tilde{\rho}(z)}{z^2}\sqrt{z^2-x^2}dz
    \end{equation}
    
    \item{Averaged projected velocity dispersion inside R:}
    \begin{equation}
        \sigma_{obs}^2=\langle\sigma_P^2\rangle(R)=\frac{4 \left(1-f^2\right) \sigma_{mod}^2}{\pi \sqrt{q}} \tilde{\langle\sigma_P^2\rangle}(x) \;  \textnormal{ where } \; \tilde{\langle\sigma_P^2\rangle}(x)=\frac{1}{\tilde{M}_{2D}(x)}\int_0^x\tilde{\sigma_P^2}(x^{\prime})\tilde{\Sigma}(x^{\prime})x^{\prime}dx^{\prime}
    \end{equation}
   
\end{itemize}

The dimensionless variables are $x \equiv R \sqrt{q}/r_{trun}$, $z \equiv r \sqrt{q}/r_{trun}$, $f \equiv r_{core}/r_{trun}$, and $q=1-e$ is the axis ratio. In our case, $f=0$ and the slit width employed to measure the velocity dispersion was 0\farcs4 \citep{2008PASJ...60L..27I} so we used the average dispersion inside R=0\farcs2. Taking into account the central values and errors for the ellipticity and the truncation radius, we obtain $\sigma_{mod,prior}=325\pm20$ km s$^-1$ as the prior for the model velocity dispersion.

\end{document}